\documentclass[runningheads]{llncs}
\usepackage{graphicx}
\usepackage{comment}
\usepackage{xcolor}
\usepackage{graphics}
\usepackage{graphicx}
\usepackage[font=bf]{caption}
\usepackage{hhline}
\usepackage{bigstrut}

\usepackage{colortbl}
\usepackage[utf8]{inputenc}

\begin{document}
\title{Easing Automatic Neurorehabilitation via Classification and Smoothness Analysis}
\titlerunning{Easing Automatic Neurorehabilitation}
%
%
\author{First Author\inst{1}\orcidID{0000-1111-2222-3333} \and
Second Author\inst{2,3}\orcidID{1111-2222-3333-4444} \and
Third Author\inst{3}\orcidID{2222--3333-4444-5555}}
\author{Asma Bensalah\inst{1}\orcidID{0000-0002-2405-9811} \and
Alicia Forn{\'e}s\inst{1}\orcidID{0000-0002-9692-5336} \and
Cristina Carmona-Duarte\inst{2}\orcidID{0000-0002-4441-6652} \and Josep Llad{\'o}s\inst{1}\orcidID{0000-0002-4533-4739} }

\authorrunning{A. Bensalah et al.}
%
\institute{Computer Vision Center, Computer Science Department, \\Universitat Aut\`{o}noma de Barcelona, Spain\\
\email{\{abensalah, afornes, josep\}@cvc.uab.es}\\
 \and
Universidad de Las Palmas de Gran Canaria, Spain\\
\email{\{cristina.carmona\}@ulpgc.es}}
\maketitle              
\begin{abstract}

Assessing the quality of movements for post-stroke patients during the rehabilitation phase is vital given that there is no standard stroke rehabilitation plan for all the patients. In fact, it depends basically on the patient's functional independence and its progress along the rehabilitation sessions.
To tackle this challenge and make neurorehabilitation more agile, we propose an automatic assessment pipeline that starts by recognising patients' movements by means of a shallow deep learning architecture, then measuring the movement quality  using jerk measure and related measures.
A particularity of this work is that the dataset used is clinically relevant, since it represents movements inspired from Fugl-Meyer a well common upper-limb clinical stroke assessment scale for stroke patients. We show that it is possible to detect the contrast between healthy and patients movements in terms of smoothness, besides achieving conclusions about the patients' progress during the rehabilitation sessions that correspond to the clinicians' findings about each case.


\end{abstract}
\section{Introduction}
Neurological disorders result in cognitive and motor impairments. The stroke survivors in particular may face deficits in motor functions in one side of the body.
These function deficits are addressed through rehabilitation sessions to partially or fully recover the functional independence of the patient~\cite{rehab}. One of the central challenges, during this phase, is the assessment of the patient's evolution. Essentially, notable progress in post-stroke patient cases happens during the first weeks namely the critical windows of heightened neuroplasticity~\cite{window}. After that, the non-linear recovery function reaches asymptotic levels. For all above reasons, both timing and treatment intensity in that critical period of time should be optimised. Thus it is indispensable to monitor patient's progress continuously and accurately, in order to maximise the patient's recovery by the end of the critical window.   
For long years, the way to proceed has been to use specific clinical scales~\cite{scale}. In practice, the patients' motor functions are evaluated once or twice in ten days. Ergo, the drawback of such an approach is that the patient's evolution is not assessed whenever patient is out of the rehabilitation room. In fact, its the patient's daily activities performance that best reflect his functional independence.

One way to cope with this limitation is to automatize the assessment in order to help clinicians to asses efficiently the patient. Many issues arise when automatizing: firstly, determining the movement nature throughout a continuous recording for hours; secondly, finding out which measures describe best the movement quality.


To address the previous mentioned issues, we propose a framework to automatically assess patients' movements. The framework has two parts:
    \begin{itemize}
        \item The first part consists of movements' classification via a shallow deep learning architecture into four key movements classes;
        \item The second part is an assessment module based on the jerk measure to ascertain the contrast between patients and healthy individuals' signals, as well as estimating the patients' evolution along the different sessions. Contrary to other existing kinematic algorithms that need more memory space and computational resources due to the number of kinematic parameters~\cite{parameters}, jerk is easier to implement in an embedded device.
    \end{itemize}



Along the rest of this paper, we give an overview of related works in Sec~\ref{related}, then we describe our classification deep learning architecture in Sec~\ref{Movement}. In Sec~\ref{movement_smoothness}, we give an overview of movement smoothness measurements. Next, we describe our setup in Sec~\ref{setup}. Then, we present our results and findings in Sec~\ref{results}. 

\section{Related work} \label{related}
Spotting a sequence in a signal aims to retrieve the signal or parts of it that are relevant for a given query. Depending on the nature of the query, many sequence spotting tasks arise~\cite{kws-audio}~\cite{kws-image}~\cite{spot-facial}. If the signal is a series of one or many different modalities and the query is an action, activity, motion or gesture, then we're addressing a Human Activity Recognition (HAR) task.

HAR has benefited greatly from the deep learning boom. HAR has been performed using different modalities: RGB images~\cite{rgb}, skeleton~\cite{skeleton}, acceleration~\cite{acceleration}, wifi~\cite{wifi}... 

Acceleration is a broadly exploited  modality for action recognition due to the fact that it is an non-invasive sensing method thus there are no privacy constraint issues. HAR through acceleration is possible because often humans perform a movement in the same qualitative way~\cite{review}.
HAR is either performed using traditional learning algorithms, for instance, support vector machines~\cite{svm}, k-nearest neighbors~\cite{knn} or employing deep learning models such as Convolutional Neural Networks (CNN)~\cite{cnn}, Recurrent Neural Networks(RNNs)~\cite{rnn} or Long short-term memory (LSTM)~\cite{lstm}.
According to~\cite{review-har}, 22\% of the HAR works were dedicated to health applications.

On the other hand, to assess recognized movements particularly for stroke patients there is no general agreement on how to obtain a movement smoothness indicator or what measure describes it best~\cite{phd}. One reason for that is the vague understanding of the neurophysiology behind movements' quality, as it is the case for upper limb movements~\cite{Kordelaar}. 
According to~\cite{nl_phd}, works about smoothness measures for stroke patients fall mainly in five different categories: trajectory related metrics~\cite{curvature}, velocity related metrics~\cite{velocity_1}~\cite{velocity_2}, acceleration related metrics~\cite{acceleration_1}~\cite{acceleration_2}, jerk related metrics~\cite{jerk_1} and other metrics~\cite{other}.
As explained above, HAR tasks have been tackled in many ways, as well as the assessment movement quality question. 
In our work, and given the few available data, we opt for a shallow deep learning architecture for HAR; moreover, we explore the use of jerk measure for assessing patients quality movements.

\section{Movement Classification} \label{Movement}
We got inspired from Supratak's model~\cite{tiny} that was designed to tackle the Polysomnography challenges, which is one of the ways to assess sleep quality~\cite{poly}. Traditionally, Polysomnography is performed by a group of experts that annotate recorded data, using a sleep stage scoring. 

To alleviate the previous limitations the architecture implements a data augmentation module and less signal processing steps in the pipeline.

\subsection{Architecture}

\begin{figure}[ht!]
\centering
\includegraphics[width=\textwidth]{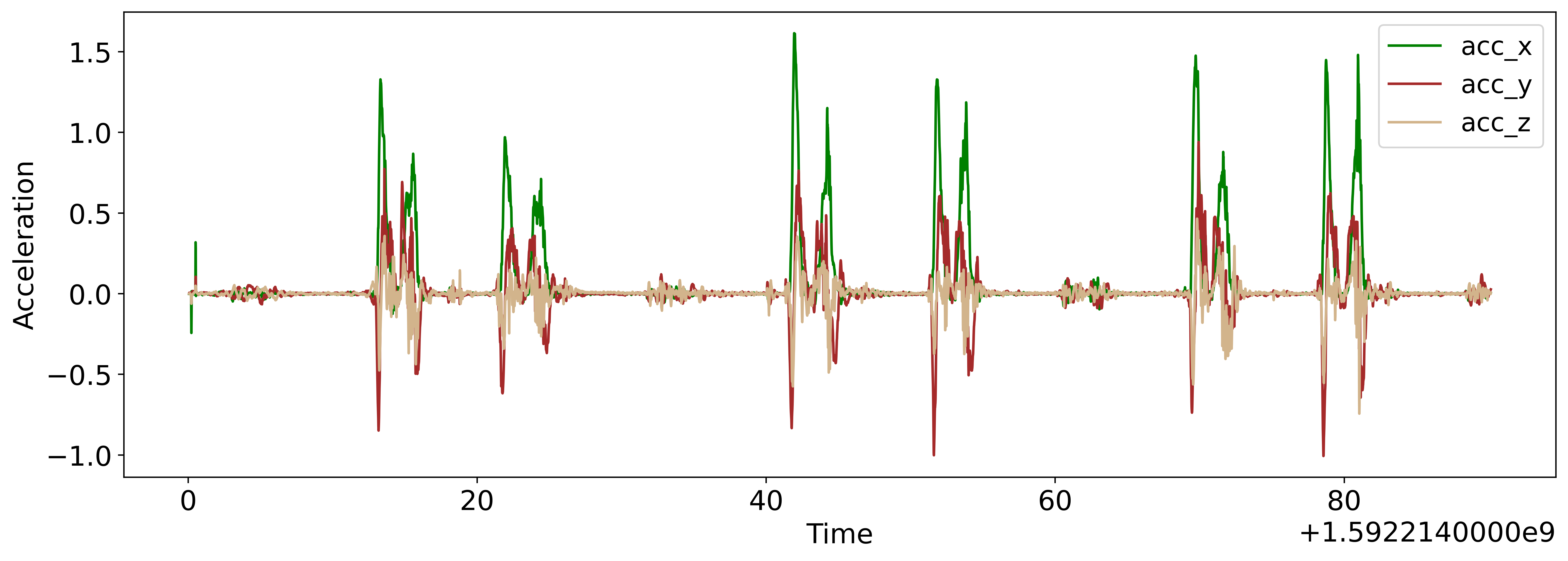}
\caption{Raw acceleration signal of a healthy individual.}
\label{fig:Healthy-acceleration-repetition}
\end{figure}

\begin{figure}[ht!]
\centering
\includegraphics[width=\textwidth]{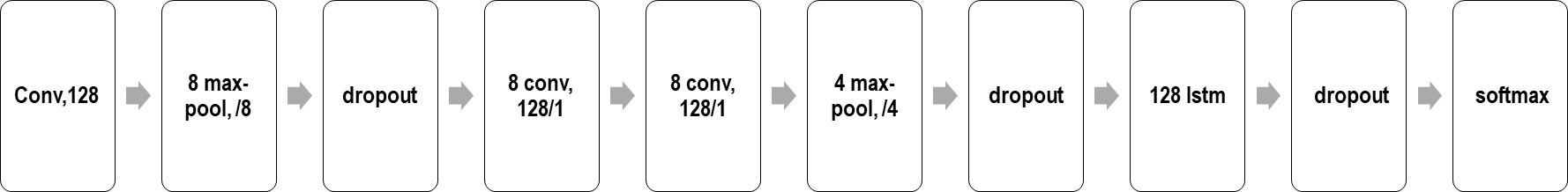}
\caption{Model architecture.}
\label{fig_cnn}
\end{figure}

The model is fed with  epochs of our raw acceleration signal (see Figure~\ref{fig:Healthy-acceleration-repetition}). We classify the movements into one of the four key movement classes: M1, M2, M3, M4.
The first component of the model is composed of four CNN layers with the aim of extracting time-invariant features from the raw signal. A max-pool and a dropout layers are introduced after the first CNN layer and the last one, as exemplified in Figure~\ref{fig_cnn}.
The second component is designed to learn temporal dependencies of the the raw signal (sequence learning). This is done via one LSTM layer followed by a dropout layout, and together they form a unidirectional RNN. The unidirectional RNN is supposed to learn time transition rules. The unidirectionality of the LSTM results in eliminating the forward pass, hence, reducing the number of hyperparameters and the computational resources.

Since our dataset is balanced, the weighted cross-entropy loss is set to 1 for all classes. Furthermore, to address the scarce data issue, a data augmentation is performed on the original data, every training epoch. Data augmentation is carried out by shifting the signal through the time axis, the shifting span is from a certain range of the epoch duration.
The model is pretrained with the Sleep-EDF dataset~\cite{sleep-edf}. 

\section{Movement smoothness} \label{movement_smoothness}
Following the description of the classification architecture, we present the movements' smoothness measurements next. 
Quantifying a movement quality can be performed in many ways. Measuring the position relative to time, is one of them. Velocity \overrightarrow{v} (equation~\ref{eq:velocity}), acceleration \overrightarrow{a} (equation~\ref{eq:acceleration}), jerk \overrightarrow{j} (equation~\ref{eq:jerk}) and snap \overrightarrow{s} (equation~\ref{eq:snap}) are respectively the first, second, third and fourth derivative of the trajectory \overrightarrow{x} with respect to time, are the most widespread used measuring quantities~\cite{calculating}.
Those are the same measures used by the human body to manage its balance. More specifically, this is handled by the sensorial functions of the vestibular system that provides information such as body position together with gravity direction~\cite{ear-balance}.
If an object is in motion, it experiences velocity. When velocity is not constant, the object is said to have an acceleration which is not equal to zero. If acceleration is varying over time, then emerges a sensation of jerkiness of the movement. 
Since attention was brought back to jerk in~\cite{schot}, it has had many applications in the science and technology fields~\cite{jerk}~\cite{jerk1}~\cite{jerk2}. 
Jerk should always take into account when vibrations occur, also whenever an abrupt transition happens~\cite{jerk}. For example, jerk is considered when designing railways to ensure a smooth motion whenever train changes from a straight line to a curved one, equally when ensuring that an industrial tool fails too soon because of fast acceleration changes.

When analysing a human movement by looking at its acceleration, it is axis orientation dependent. A small rotation of the wrist while recording data can result in a lot of noise in an axis acceleration. Hence, in this work, we focus on jerk as a movement quality measure, in particular, as a smoothness indicator. 
Ultimately, jerk is easier to implement in a an embedded device, unlike other existing kinematic algorithms that need more space due to the number of kinematic parameters.


\begin{equation}\label{eq:velocity}
\overrightarrow{v}(t) = \frac{\mathrm{d}\overrightarrow{x}(t)}{\mathrm{d}t}
\end{equation}
\begin{equation}\label{eq:acceleration}
\overrightarrow{a}(t) = \frac{\mathrm{d}\overrightarrow{v}(t)}{\mathrm{d}t}
\end{equation}
\begin{equation}\label{eq:jerk}
\overrightarrow{j}(t) = \frac{\mathrm{d}\overrightarrow{a}(t)}{\mathrm{d}t}
\end{equation}
\begin{equation}\label{eq:snap}
\overrightarrow{s}(t) = \frac{\mathrm{d}\overrightarrow{j}(t)}{\mathrm{d}t}
\end{equation}

\section{Setup}\label{setup}

\subsection{Dataset}
The dataset used was recorded as a part of  \emph{3D kinematics for remote patient monitoring} (RPM3D) project\footnote{\url{http://dag.cvc.uab.es/patientmonitoring/}}, aiming to build an automatic pipeline for stroke patients. A dataset for stroke patients and healthy subjects along with a classification baseline was published~\cite{icpr}.
Patients and healthy individuals were given a smartwatch in each hand. Healthy individuals were recorded once while patients were recorded during four different sessions. The time interval between patients' sessions is between one or two weeks.
Initially, to assess a stroke patient upper limb motor functions, an assessment is performed once in a week or ten days. The best-known scale to asses sensorimotor impairments within stroke patients is the Fugl-Meyer Assessment~\cite{fugl}. For this reason, authors were inspired from the Fugl-Meyer movements to design their set of key movements $\mathcal{M}_i, i\in \left [ 1,4 \right ]$, thusly:
\begin{itemize}
    \item Movement $\mathcal{M}_1$: shoulder extension/flexion.
    \item Movement $\mathcal{M}_2$: shoulder abduction.
    \item Movement $\mathcal{M}_3$: external/internal shoulder rotation.
    \item Movement $\mathcal{M}_4$: elbow flexion/extension
\end{itemize} 
\subsubsection{Scenarios}
The  experiments were held into two different setups: a constrained scenario L1 and unconstrained one L2. These are described as follows:
\begin{itemize}
    \item Scenario L1: it represents a constrained scenario, where individuals perform four key movements $\mathcal{M}_i$: once with the dominant hand, second using the non-dominant one and lastly with both hands.
    \item Scenario L2: it represents the unconstrained scenario, composed of a sequence of key movements $\mathcal{M}_i$ along with a set of other non-target movements $\mathcal{R}_j$, $j\in \left [ 1,19 \right ]$. $\mathcal{R}_i$ movements are a list of usual daily activities such as: drinking, setting on a chair,... 
    The movements are carried out in a random order.
\end{itemize}

\subsection{Pipeline}
\begin{figure}[ht!]
\centering
\includegraphics[width=\textwidth]{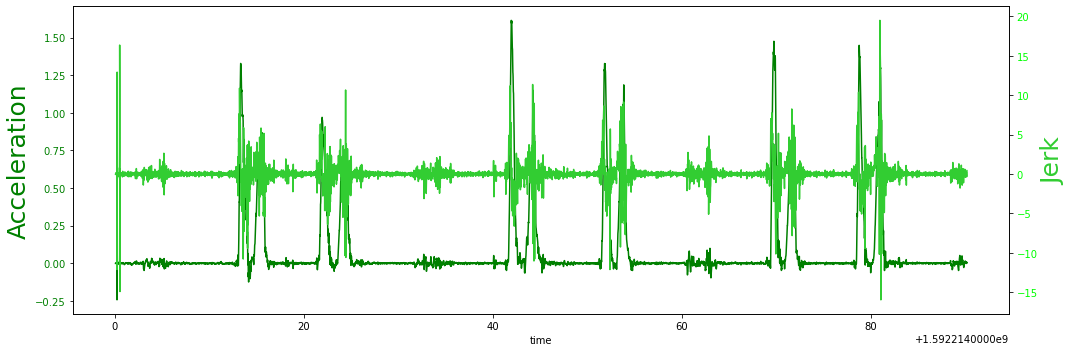}
\caption{Acceleration and jerk for a repetition of movements - Healthy individual.}
\label{fig:Healthy-Jerk-repetition}
\end{figure}

We start by classifying movements into the four main classes: M1, M2, M3, M4 (shoulder extension/flexion, shoulder abduction, external/internal shoulder rotation, elbow flexion/extension).Then we compute the jerk value for a signal that represents performing one movement, for several times, such as shown in Figure~\ref{fig:Healthy-Jerk-repetition} 
to inspect the global acceleration patterns of a movement.

After that, we compute the jerk for a smaller fragment of the previous signal (one well segmented movement), for a more accurate smoothness estimation.

\section{Results}\label{results}
Results below are  related to the classification of L1 movements and their smoothness analysis.
\subsection{Classification}
For the experiments, the data is divided into 80\% for training, and 20\% for testing. The testing accuracy reaches an average of~77,01\%. We experience a decrease of accuracy in some epochs, which we believe is due to the small size of the training data set.

\subsection{Smoothness}
\begin{figure}[ht!]
\centering
\includegraphics[width=\textwidth]{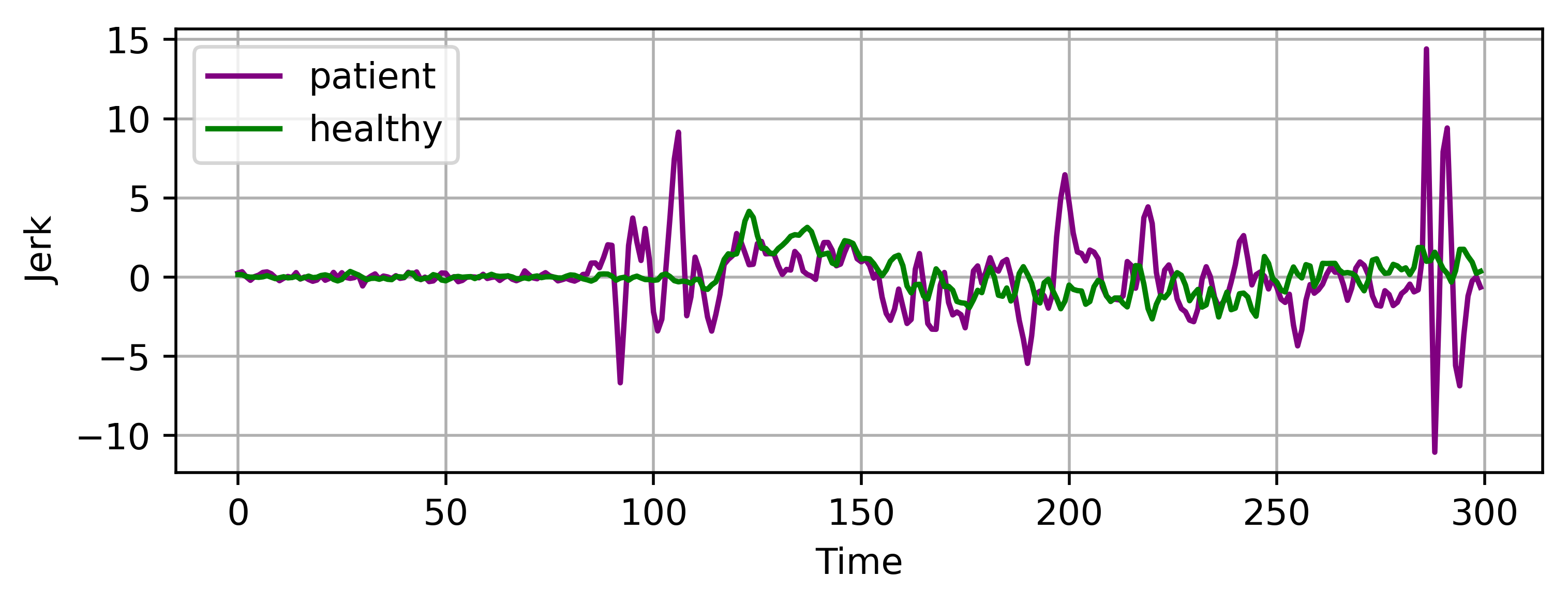}
\caption{Jerk values for a patient and healthy individual performing the same movement.}
\label{fig:Healthy_patient_jerk}
\end{figure}

In Table \ref{tab:jerk_x} we give information about the jerk values for well segmented movements of healthy and patient individuals, along axis x.
Theoretically, the jerk value should be lower in the case of healthy individuals compared to patients, because their ability to move and perform the movements in a smoother way is superior, thus their movements are less jerky (see Figure~\ref{fig:Healthy_patient_jerk})~\cite{nl_phd}.

\begin{figure}[ht!]
\centering
\includegraphics[width=\textwidth]{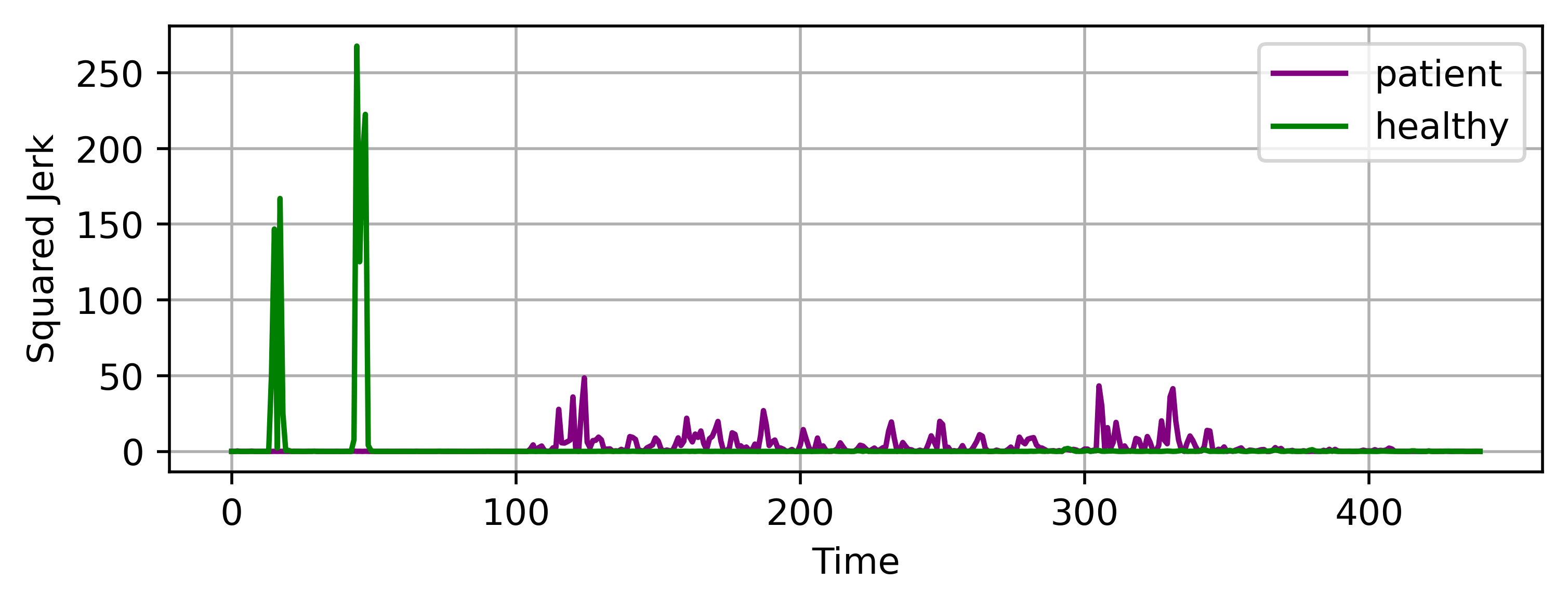}
\caption{Healthy vs patient signal.}
\label{fig:Healthy-patient-peaks}
\end{figure}

As observed in Table \ref{tab:jerk_x}, this is the case for M2, M3, M4. For instance, regarding movement M3, the absolute value of the jerk mean for patients is  0.7 times the absolute value of the jerk mean for healthy subjects. 
Simultaneously, the trend in Table \ref{tab:jerk_x} is that the maximum of jerk within the healthy population is greater than the patients' one, for all four movements M1,M2,M3, and M4. 
\newline
\begin{table}[ht!]
  \centering
  \caption{Jerk measures for patients and healthy individuals, along axis x.}
    \begin{tabular}{|r|r|r|r|r|r|r|}
\cline{2-7}    \multicolumn{1}{r|}{} & \multicolumn{6}{c|}{ Jerk Measure} \bigstrut\\
\cline{2-7}    \multicolumn{1}{r|}{} & \multicolumn{2}{c|}{Mean } & \multicolumn{2}{c|}{Max} & \multicolumn{2}{c|}{Min} \bigstrut\\
    \hline
    axis x & \multicolumn{1}{c|}{Healthy} & \multicolumn{1}{c|}{Patient} & \multicolumn{1}{c|}{Healthy} & \multicolumn{1}{c|}{Patient} & \multicolumn{1}{c|}{Healthy} & \multicolumn{1}{c|}{Patient} \bigstrut\\
    \hline
    M1    & 0,00500712 & -0,006998 & 497,99 & 145,54 & -1,96E+02 & -138,99182 \bigstrut\\
    \hline
    M2    & -0,0004051 & -0,001621 & 157,56 & 69,46 & -186,062127 & -152,652 \bigstrut\\
    \hline
    M3    & -0,0023341 & 0,0016529 & 102,74 & 84,58 & -1,27E+02 & -62,973556 \bigstrut\\
    \hline
    M4    & 0,00031141 & 0,0006176 & 161,04 & 107,14 & -1,50E+02 & -114,20477 \bigstrut\\
    \hline
    \end{tabular}%
  \label{tab:jerk_x}%
\end{table}%
The jerk represents the change in acceleration. In that sense, to gain more understanding of the movements' smoothness, we went for jerk related measures, which are calculated based on the absolute value of the jerk. In this work, we focus on the squared jerk measure.
Table \ref{tab:squared_jerk_x} shows the mean, maximum and minimum values of the squared jerk measure. Notice that the trend in Table \ref{tab:jerk_x} is that the mean jerk within healthy individuals is lower, which corresponds to the theory premises' that the jerkier and less smooth the movement is, the higher is the jerk value. Hence, patients should have higher jerk values. 
Nonetheless, this is not the case for the squared mean jerk. The general pattern in Table \ref{tab:squared_jerk_x} is that the healthy population's squared jerk mean is higher than patients. We think that this could be related to the fact that a patient signal is noisier than healthy individual one because patients are slower, thus a patient's signal has more peaks and more cumulative noise (see Figure~\ref{fig:Healthy-patient-peaks}).

\begin{table}[ht!]
  \centering
  \caption{Squared jerk measures for patients and healthy individuals, along axis x.}
    \begin{tabular}{|l|r|r|r|r|r|r|}
\cline{2-7}    \multicolumn{1}{r|}{} & \multicolumn{6}{c|}{Suquared Jerk Measure} \bigstrut\\
\cline{2-7}    \multicolumn{1}{r|}{} & \multicolumn{2}{c|}{Mean } & \multicolumn{2}{c|}{Max} & \multicolumn{2}{c|}{Min} \bigstrut\\
    \hline
    \multicolumn{1}{|c|}{axis x} & \multicolumn{1}{c|}{Healthy} & \multicolumn{1}{c|}{Patient} & \multicolumn{1}{c|}{Healthy} & \multicolumn{1}{c|}{Patient} & \multicolumn{1}{c|}{Healthy} & \multicolumn{1}{c|}{Patient} \bigstrut\\
    \hline
    M1    & 19,96 & 7,65  & 247993,61 & 21182,96 & 8,93E-12 & 2,22E-12 \bigstrut\\
    \hline
    M2    & 18,11 & 5,76  & 34619,12 & 23302,63 & 0     & 8,42E-10 \bigstrut\\
    \hline
    M3    & 14,40 & 3,75  & 16056,08 & 7153,85 & 3,55E-11 & 0 \bigstrut\\
    \hline
    M4    & 26,19 & 4,48  & 25934,96 & 13042,73 & 8,93E-12 & 0 \bigstrut\\
    \hline
    \end{tabular}%
  \label{tab:squared_jerk_x}%
\end{table}%

\begin{figure}[ht!]
\centering
\includegraphics[width=\textwidth]{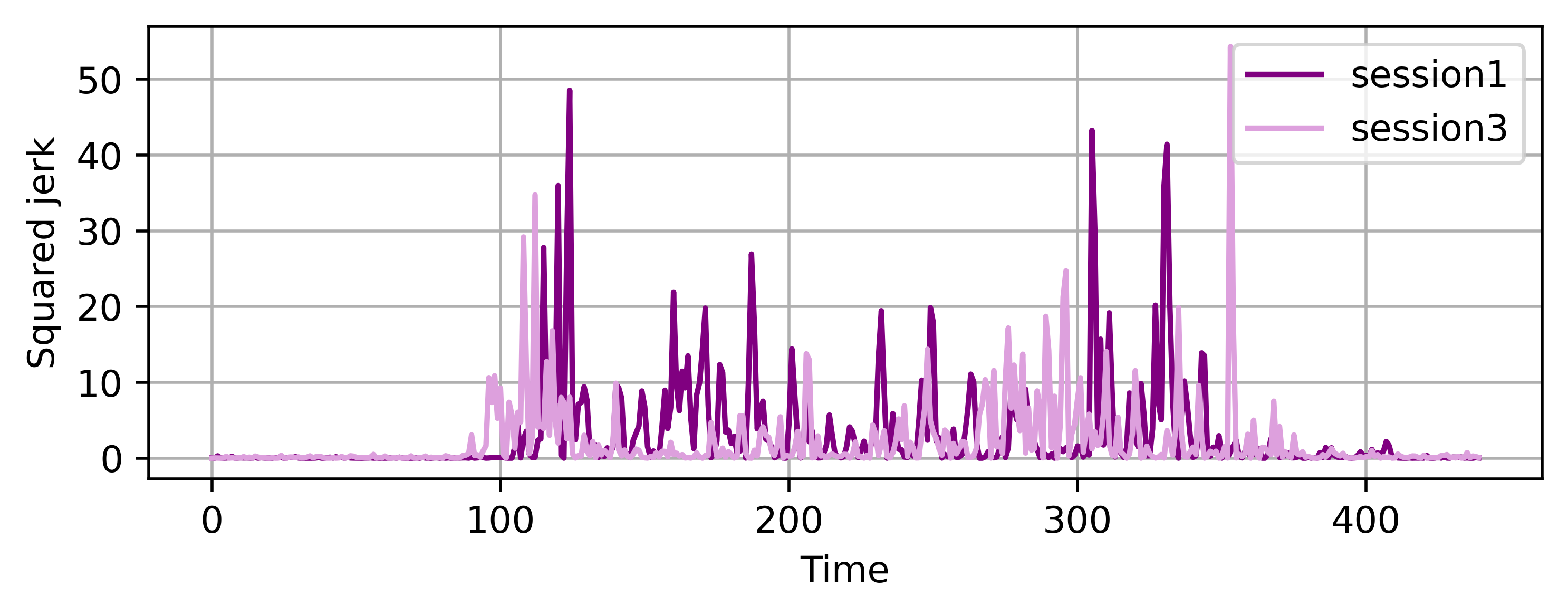}
\caption{Squared jerk values for movement M1 in session 1 and 3- Patient 100.}
\label{fig:squared_jerk_100_m1}
\end{figure}

Tables~\ref{tab:patient100},~\ref{tab:patient101},~\ref{tab:patient102},~\ref{tab:patient103} provide information about the squared jerk measure for four patients, namely: 100, 101, 102, 103. It indicates the evolution of four patients through four sessions, along axis x.
Table~\ref{tab:patient100} gives information about the squared jerk measures: mean, maximum and minimum for patient 100. Patient's performance for movement M1 is better in sessions 3 and 4. Figure~\ref{fig:squared_jerk_100_m1} shows a less jerky M1 in session 3 compared to the first session. At the same time patient 100 reaches the most significant improvement for M3 and M4 in the third session. Yet, movement M2 squared jerk mean values present no improvement during the four sessions.

\begin{table}[ht!]
  \centering
  \caption{Squared jerk measures for patient 100 across four sessions, along axis x.}
   \resizebox{\columnwidth}{!}{ \begin{tabular}{|l|r|r|r|r|r|r|r|r|r|r|r|r|}
\cline{2-13}    \multicolumn{1}{r|}{} & \multicolumn{12}{c|}{Patient 100} \bigstrut\\
\cline{2-13}    \multicolumn{1}{r|}{} & \multicolumn{4}{c|}{Mean }    & \multicolumn{4}{c|}{Max}      & \multicolumn{4}{c|}{Min} \bigstrut\\
    \hline
    \multicolumn{1}{|c|}{axis x} & \multicolumn{1}{l|}{Session 1} & \multicolumn{1}{l|}{Session2} & \multicolumn{1}{c|}{Session3} & \multicolumn{1}{c|}{Session4} & \multicolumn{1}{l|}{Session 1} & \multicolumn{1}{l|}{Session2} & \multicolumn{1}{c|}{Session3} & \multicolumn{1}{c|}{Session4} & \multicolumn{1}{l|}{Session 1} & \multicolumn{1}{l|}{Session2} & \multicolumn{1}{c|}{Session3} & \multicolumn{1}{c|}{Session4} \bigstrut\\
    \hline
    M1    & \cellcolor[rgb]{ 1,  1,  0}12,88 & 13,04 & \cellcolor[rgb]{ 1,  .753,  0}5,09 & \cellcolor[rgb]{ 1,  .753,  0}10,16 & 18882,36 & 12129,74 & 4166,80 & 21182,96 & 8,38E-09 & 1,67E-08 & 1,75E-07 & 2,30E-09 \bigstrut\\
    \hline
    M2    & \cellcolor[rgb]{ 1,  1,  0}4,17 & 9,07  & 12,74 & 19,92 & 715,66 & 1246,53 & 23302,63 & 7364,52 & 9,35E-08 & 4,49E-08 & 8,42E-10 & 1,88E-08 \bigstrut\\
    \hline
    M3    & \cellcolor[rgb]{ 1,  1,  0}9,16 & \cellcolor[rgb]{ 1,  .753,  0}6,25 & \cellcolor[rgb]{ 1,  .753,  0}4,65 & 8,28  & 3738,07 & 2385,56 & 976,07 & 7153,85 & 8,53E-10 & 2,05E-10 & 5,18E-09 & 3,94E-08 \bigstrut\\
    \hline
    M4    & \cellcolor[rgb]{ 1,  1,  0}6,28 & 30,45 & \cellcolor[rgb]{ 1,  .753,  0}4,80 & 7,14  & 1804,63 & 13042,73 & 1246,78 & 670,45 & 1,98E-08 & 1,28E-08 & 1,20E-08 & 0 \bigstrut\\
    \hline
    \end{tabular}}%
  \label{tab:patient100}%
\end{table}%

\begin{table}[ht!]
  \centering
  \caption{Squared jerk measures for patient 101 across four sessions, along axis x.}
   \resizebox{\columnwidth}{!}{\begin{tabular}{|l|r|r|r|r|r|r|r|r|r|r|r|r|}
\cline{2-13}    \multicolumn{1}{r|}{} & \multicolumn{12}{c|}{Patient 101} \bigstrut\\
\cline{2-13}    \multicolumn{1}{r|}{} & \multicolumn{4}{c|}{Mean }    & \multicolumn{4}{c|}{Max}      & \multicolumn{4}{c|}{Min} \bigstrut\\
    \hline
    axis x & \multicolumn{1}{l|}{Session 1} & \multicolumn{1}{l|}{Session2} & \multicolumn{1}{c|}{Session3} & \multicolumn{1}{c|}{Session4} & \multicolumn{1}{l|}{Session 1} & \multicolumn{1}{l|}{Session2} & \multicolumn{1}{c|}{Session3} & \multicolumn{1}{c|}{Session4} & \multicolumn{1}{l|}{Session 1} & \multicolumn{1}{l|}{Session2} & \multicolumn{1}{c|}{Session3} & \multicolumn{1}{c|}{Session4} \bigstrut\\
    \hline
    M1    & \cellcolor[rgb]{ 1,  1,  0}7,61 & \cellcolor[rgb]{ 1,  .753,  0}2,62 & \cellcolor[rgb]{ 1,  .753,  0}4,63 & \cellcolor[rgb]{ 1,  .753,  0}5,23 & 1357,01 & 445,53 & 581,59 & 968,97 & 1,64E-07 & 2,22E-12 & 4,59E-08 & 3,55E-11 \bigstrut\\
    \hline
    M2    & \cellcolor[rgb]{ 1,  1,  0}2,03 & 2,16  & 7,52  & 5,20  & 267,21 & 484,64 & 3781,94 & 1086,66 & 1,04E-07 & 1,05E-08 & 2,22E-08 & 2,82E-09 \bigstrut\\
    \hline
    M3    & \cellcolor[rgb]{ 1,  1,  0}2,14 & 4,05  & 3,21  & 8,03  & 406,89 & 2257,01 & 604,45 & 3965,67 & 6,13E-08 & 8,39E-09 & 4,69E-09 & 3,60E-08 \bigstrut\\
    \hline
    M4    & \cellcolor[rgb]{ 1,  1,  0}0,58 & 0,73  & 2,88  & 2,49  & 44,56 & 163,14 & 1068,82 & 1034,61 & 2,24E-09 & 1,52E-08 & 8,85E-10 & 5,00E-10 \bigstrut\\
    \hline
    \end{tabular}}%
  \label{tab:patient101}%
\end{table}%

\begin{figure}[ht!]
\centering
\includegraphics[width=\textwidth]{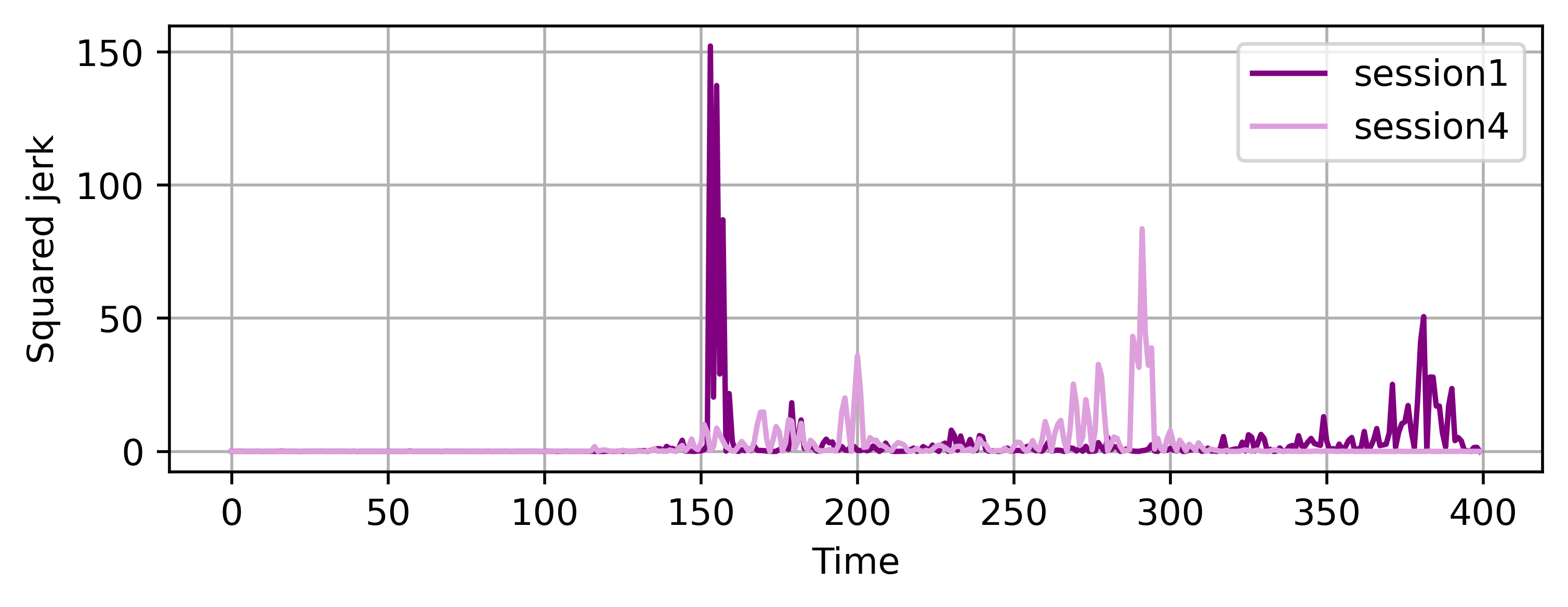}
\caption{Squared jerk values for movement M4 in session 1 and 4- Patient 101.}
\label{fig:squared_jerk_101_m4}
\end{figure}

As for Patient 101 (see Table~\ref{tab:patient101}), the mean squared jerk values have increased during the four rehabilitation sessions, as illustrated in Figure~\ref{fig:squared_jerk_101_m4}. The Figure shows a less smooth M4 movement in the last session, except for movement M1, which experiences a decrease in the mean squared jerk value compared to the first session.

Table~\ref{tab:patient102} depicts patient's 102 data, in which movements M1, M2, M3 witness a gradual decrease of squared jerk mean until the last rehabilitation session. Contrary, the M1 mean squared jerk stops lessening after the second session.

It is clear that the three patients 100, 102, 103 have reached lower squared jerk means than those of their first sessions, for at least three movements. 

\textbf{How good is jerk as a smooth indicator?} Overall, it is not trivial to compare the jerk values of healthy individuals to the ones of the patients owing to the way the patients performed the movements. In particular, when patients have difficulties to perform the movements in a consistent way, it implies that a simple comparison of healthy movements' jerk values and patients ones is not always conclusive. For example, in the case of movement M1, the jerk mean value is higher within healthy samples than the patient samples. 
Despite that, the mean squared jerk values provide interesting insights concerning the evolution of patients across the four sessions. Our conclusions do align with the clinicians' closures: most patients' smoothness improved when compared to the first session. Additionally, for patient 101, the patient that presented more motor function issues during the sessions, we observed the least improvement in terms of squared jerk mean values.
\begin{table}[ht!]
  \centering
  \caption{squared jerk measures for patient 102 across four sessions, along axis x.}
    \resizebox{\columnwidth}{!}{\begin{tabular}{|l|r|r|r|r|r|r|r|r|r|r|r|r|}
\cline{2-13}    \multicolumn{1}{r|}{} & \multicolumn{12}{c|}{Patient 102} \bigstrut\\
\cline{2-13}    \multicolumn{1}{r|}{} & \multicolumn{4}{c|}{Mean }    & \multicolumn{4}{c|}{Max}      & \multicolumn{4}{c|}{Min} \bigstrut\\
    \hline
    axis x & \multicolumn{1}{l|}{Session 1} & \multicolumn{1}{l|}{Session2} & \multicolumn{1}{c|}{Session3} & \multicolumn{1}{c|}{Session4} & \multicolumn{1}{l|}{Session 1} & \multicolumn{1}{l|}{Session2} & \multicolumn{1}{c|}{Session3} & \multicolumn{1}{c|}{Session4} & \multicolumn{1}{l|}{Session 1} & \multicolumn{1}{l|}{Session2} & \multicolumn{1}{c|}{Session3} & \multicolumn{1}{c|}{Session4} \bigstrut\\
    \hline
    M1    & \cellcolor[rgb]{ 1,  1,  0}4,27 & \cellcolor[rgb]{ 1,  .753,  0}1,18 & 4,82  & 7,09  & 5256,19 & 49,80 & 3829,05 & 7841,56 & 1,61E-08 & 2,90E-09 & 1,50E-08 & 8,84E-12 \bigstrut\\
    \hline
    M2    & \cellcolor[rgb]{ 1,  1,  0}2,89 & 4,61  & \cellcolor[rgb]{ 1,  .753,  0}2,21 & \cellcolor[rgb]{ 1,  .753,  0}1,78 & 473,36 & 2788,06 & 353,56 & 739,78 & 5,98E-09 & 4,09E-08 & 8,61E-10 & 3,77E-08 \bigstrut\\
    \hline
    M3    & \cellcolor[rgb]{ 1,  1,  0}1,03 & 1,56  & \cellcolor[rgb]{ 1,  .753,  0}1,03 & \cellcolor[rgb]{ 1,  .753,  0}0,58 & 88,71 & 405,24 & 117,23 & 81,77 & 1,26E-09 & 0     & 3,21E-08 & 6,56E-09 \bigstrut\\
    \hline
    M4    & \cellcolor[rgb]{ 1,  1,  0}1,90 & \cellcolor[rgb]{ 1,  .753,  0}1,25 & \cellcolor[rgb]{ 1,  .753,  0}1,67 & \cellcolor[rgb]{ 1,  .753,  0}0,84 & 306,19 & 165,53 & 184,44 & 125,36 & 1,22E-08 & 1,77E-09 & 3,01E-08 & 2,59E-09 \bigstrut\\
    \hline
    \end{tabular}}%
  \label{tab:patient102}%
\end{table}%

\begin{table}[ht!]
  \centering
  \caption{squared jerk measures for patient 103 across four sessions, along axis x.}
    \resizebox{\columnwidth}{!}{\begin{tabular}{|l|r|r|r|r|r|r|r|r|r|r|r|r|}
\cline{2-13}    \multicolumn{1}{r|}{} & \multicolumn{12}{c|}{Patient 103} \bigstrut\\
\cline{2-13}    \multicolumn{1}{r|}{} & \multicolumn{4}{c|}{Mean }    & \multicolumn{4}{c|}{Max}      & \multicolumn{4}{c|}{Min} \bigstrut\\
    \hline
    \multicolumn{1}{|c|}{axis x} & \multicolumn{1}{l|}{Session 1} & \multicolumn{1}{l|}{Session2} & \multicolumn{1}{c|}{Session3} & \multicolumn{1}{c|}{Session4} & \multicolumn{1}{l|}{Session 1} & \multicolumn{1}{l|}{Session2} & \multicolumn{1}{c|}{Session3} & \multicolumn{1}{c|}{Session4} & \multicolumn{1}{l|}{Session 1} & \multicolumn{1}{l|}{Session2} & \multicolumn{1}{c|}{Session3} & \multicolumn{1}{c|}{Session4} \bigstrut\\
    \hline
    M1    & \cellcolor[rgb]{ 1,  1,  0}8,57 & 16,17 & 9,97  & 9,47  & 2570,08 & 19318,73 & 2733,30 & 1991,19 & 2,35E-08 & 2,79E-08 & 5,36E-11 & 7,71E-08 \bigstrut\\
    \hline
    M2    & \cellcolor[rgb]{ 1,  1,  0}2,80 & 4,34  & \cellcolor[rgb]{ 1,  .753,  0}1,88 & 3,79  & 500,42 & 2143,84 & 222,50 & 1488,18 & 2,18E-07 & 1,73E-09 & 1,29E-09 & 4,98E-08 \bigstrut\\
    \hline
    M3    & \cellcolor[rgb]{ 1,  1,  0}1,66 & 2,14  & 1,86  & \cellcolor[rgb]{ 1,  .753,  0}1,34 & 1039,28 & 707,50 & 297,98 & 331,72 & 8,53E-09 & 2,33E-10 & 2,86E-08 & 3,82E-08 \bigstrut\\
    \hline
    M4    & \cellcolor[rgb]{ 1,  1,  0}3,18 & 4,12  & \cellcolor[rgb]{ 1,  .753,  0}2,11 & \cellcolor[rgb]{ 1,  .753,  0}2,52 & 4304,22 & 2560,54 & 1628,78 & 713,06 & 4,45E-10 & 8,66E-09 & 3,60E-08 & 4,36E-01 \bigstrut\\
    \hline
    \end{tabular}}%
  \label{tab:patient103}%
\end{table}%

\section{Conclusion}\label{Conclusion}
In this paper, we have presented a fully automatic assessment stroke patients pipeline, combining a deep learning model and a smoothness quality module based on the jerk measure, which is computed on movements inspired from the valid clinical functional Fugl-Meyer scale.
The classification of movements reached a good accuracy even though the dataset is small, probably due to the data augmentation performed on the original signal. The jerk has proved to be a promising measure to assess stroke patients when compared to healthy subjects, while squared jerk gives a good indication for intersession patient's performance variability. 

Alike all vision and machine learning tasks that are not image or NLP related, the data available for our task is few. Hence, in the future work will be directed toward enhancing available data and exploiting more robust smoothness measures.
\section*{Acknowledgment}
This work has been partially supported by the Spanish project RTI2018-095645-B-C21, the CERCA Program / Generalitat de Catalunya and the FI fellowship AGAUR 2020 FI-SDUR 00497 (with the support of the Secretaria d’Universitats i Recerca of the Generalitat de Catalunya and the Fons Social Europeu).

\label{}

%
%

%
%
\bibliographystyle{unsrt}
\bibliography{bib.bib}

\end{document}